\documentclass[10pt]{amsart}
\usepackage{amssymb,amscd}
\usepackage{verbatim}

\usepackage{xcolor}

\usepackage{amsmath,amssymb,graphicx,mathrsfs}   
\usepackage{enumerate}
\usepackage[colorlinks=true,allcolors = blue]{hyperref} 

\usepackage{tikz}
\usetikzlibrary{matrix}

\usepackage[all]{xy}

\usepackage{amssymb,amsfonts,amsthm,amsmath,calligra}
\usepackage{slashed}
\usepackage{yfonts}
\usepackage{mathrsfs,pifont}
\usepackage{float}

\usepackage[all]{xy}

\usepackage{tikz}

\usepackage{graphicx}
\usepackage{xcolor}

\usepackage{amssymb,amsfonts,amsthm,amsmath}
\usepackage[all]{xy}
\usepackage{slashed}
\usepackage{yfonts}
\usepackage{mathrsfs,pifont}

\let\frak\mathfrak

\def\>{\relax\ifmmode\mskip.666667\thinmuskip\relax\else\kern.111111em\fi}
\def\<{\relax\ifmmode\mskip-.333333\thinmuskip\relax\else\kern-.0555556em\fi}
\def\vsk#1>{\vskip#1\baselineskip}
\def\vv#1>{\vadjust{\vsk#1>}\ignorespaces}
\def\vvn#1>{\vadjust{\nobreak\vsk#1>\nobreak}\ignorespaces}

  \let\ssize\scriptstyle
\let\sssize\scriptscriptstyle

\let\Medskip\medskip
\def\medskip{\par\Medskip}
\let\Bigskip\bigskip
\def\bigskip{\par\Bigskip}

\let\Maketitle\maketitle
\def\maketitle{\Maketitle\thispagestyle{empty}\let\maketitle\empty}

\theoremstyle{definition}                                  
\numberwithin{equation}{section}

\theoremstyle{definition}

\let\mc\mathcal
\let\nc\newcommand

\let\la\lambda

\let\phi\varphi

\let\der\partial

\let\geq\geqslant

\let\leq\leqslant

\let\on\operatorname
\let\bi\bibitem
\let\bs\boldsymbol

\def\C{{\mathbb C}}
\def\Z{{\mathbb Z}}

\def\F{{\mathbb F}}

\def\+#1{^{\{#1\}}}

\def\Gr{\on{Gr}}

\def\beq{\begin{equation}}
\def\eeq{\end{equation}}
\def\be{\begin{equation*}}
\def\ee{\end{equation*}}

\nc{\bea}{\begin{eqnarray*}}
\nc{\eea}{\end{eqnarray*}}
\nc{\bean}{\begin{eqnarray}}
\nc{\eean}{\end{eqnarray}}

\nc{\Il}{{\mc I_{\bs\la}}}
\nc{\bla}{{\bs\la}}
\nc{\Fla}{\F_\bla}
\nc{\tfl}{{T^*\Fla}}
\nc{\GL}{{GL_n(\C)}}
\nc{\GLC}{{GL_n(\C)\times\C^*}}

\let\sd s 

\def\ddk_#1{\kk_{#1}\<\>\frac\der{\der\<\>\kk_{#1}}}

\def\bul{\mathbin{\raise.2ex\hbox{$\sssize\bullet$}}}
\def\intt{\mathchoice
{\mathop{\raise.2ex\rlap{$\,\,\ssize\backslash$}{\intop}}\nolimits}
{\mathop{\raise.3ex\rlap{$\,\sssize\backslash$}{\intop}}\nolimits}
{\mathop{\raise.1ex\rlap{$\sssize\>\backslash$}{\intop}}\nolimits}
{\mathop{\rlap{$\sssize\<\>\backslash$}{\intop}}\nolimits}}

\let\kk q 
\let\cc c

\let\Ko K

\def\GZ/{Gelfand-Zetlin}
\def\KZ/{{\slshape KZ\/}}
\def\qKZ/{{\slshape qKZ\/}}
\def\XXX/{{\slshape XXX\/}}

\nc{\A}{{\mc A}}

\nc{\hsl}{\widehat{{\frak{sl}_2}}}

\nc{\BC}{{ \mathbb C}}
\nc{\lra}{\longrightarrow}
\nc{\CO}{{\mathcal{O}}}
\nc{\BZ}{{ \mathbb Z}}
\nc{\hfn}{\hat{\frak{n}}}
\nc\Zs{{\Z/p^s\Z}}
\nc\Zo{{\Zs[z]^0}}
\nc\gr{{\on{gr}}}

\nc\fD{{\frak D}}

\newcommand{\matC}{\mathbb{C}}
\newcommand{\matN}{\mathbb{N}}

\newcommand{\matQ}{\mathbb{Q}}

\newcommand{\matZ}{\mathbb{Z}}

\newcommand{\Stab}{\mathsf{Stab}}
\newcommand{\Rm}{\mathsf{R}}
\newcommand{\Mon}{\mathsf{Mon}}
\newcommand{\kms}{\bar{\mathsf{Z}}}
\newcommand{\ems}{\bar{\mathsf{A}}}
\newcommand{\lb}{\mathcal{L}}

\newcommand{\torA}{\mathsf{A}}
\newcommand{\torZ}{\mathsf{Z}}

\newcommand{\wc}{\mathbf{B}}
\newcommand{\Mop}{\mathbf{M}}

\newcommand{\cL}{\mathcal{L}}

\newcommand{\Pic}{\mathrm{Pic}}
\newcommand{\zamon}{\Sigma}

\usepackage{tikz}

\usetikzlibrary{decorations}
\usetikzlibrary{decorations.pathmorphing}
\usetikzlibrary{calc}

\newsavebox{\hweights}
\savebox{\hweights}{%
\begin{tikzpicture}[baseline= {($(current bounding box.base)-(0pt,-30pt)$)}]
\begin{scope}
\draw[line width=0.35mm] (0,0)-- (5,5) ;

\draw[line width=0.35mm] (-1,1)-- (4,6) ;

\draw[line width=0.35mm] (-2,2)-- (3,7) ;

\draw[line width=0.35mm] (-3,3)-- (2,8) ;

\draw[line width=0.35mm] (0,0)-- (-3,3) ;
\draw[line width=0.35mm] (1,1)-- (-2,4) ;
\draw[line width=0.35mm] (2,2)-- (-1,5) ;
\draw[line width=0.35mm] (3,3)-- (0,6) ;
\draw[line width=0.35mm] (4,4)-- (1,7) ;
\draw[line width=0.35mm] (5,5)-- (2,8) ;
\node at (0,1) { $1$};
\node at (-1,2) { $2$}; \node at (1,2) { $2$};
\node at (-2,3) { $3$};\node at (0,3) { $3$};
\node at (2,3) { $3$};
\node at (-1,4) { $4$};
\node at (1,4) { $4$};
\node at (3,4) { $4$};
\node at (0,5) { $5$};
\node at (2,5) { $5$};
\node at (4,5) { $5$};
\node at (1,6) { $6$};
\node at (3,6) { $6$};
\node at (2,7) { $7$};
\end{scope}
\end{tikzpicture}}

\begin{document}

\hrule width0pt
\vsk->

\title[Enumerative geometry via elliptic stable envelope]
{Enumerative geometry via elliptic stable envelope}

\author
[Andrey Smirnov ]
{Andrey Smirnov$^{\star}$ }

\maketitle

\begin{center}
{ Department of Mathematics, University
of North Carolina at Chapel Hill\\ Chapel Hill, NC 27599-3250, USA\/}

\end{center}

\vsk>
{\leftskip3pc \rightskip\leftskip \parindent0pt \Small
{\it Key words\/}:  Quantum difference equations, quantum K-theory, elliptic stable envelope.

\vsk.6>
{\it 2020 Mathematics Subject Classification\/}: 14G33 (11D79, 32G34, 33C05, 33E30)
\par}


{\let\thefootnote\relax
\footnotetext{\vsk-.8>\noindent
$^\star\<${\sl E\>-mail}:\enspace asmirnov@email.unc.edu
}

\begin{abstract}
Assume $X$ is a variety for which the elliptic stable envelope exists.  In this note we construct natural $q$-difference equations from the elliptic stable envelope of $X$. In examples, these equations coincide with the quantum difference equations, which give a natural $q$-deformation of the Dubrovin connection of $X$. Solutions of the quantum difference equations provide generating functions counting curves in $X$. In this way, our construction connects  curve counting and equivariant elliptic cohomology.

This is an overview paper based on the author's talk at the workshop {\it The 16th MSJ-SI: Elliptic Integrable Systems, Representation Theory 
and Hypergeometric Functions}, Tokyo 2023.

\end{abstract}


\setcounter{footnote}{0}

\section{Introduction}

Ideas surrounding the so called ``$3D$-mirror symmetry'' have been recently gaining in popularity among mathematicians. 
This symmetry has its origin in theoretical physics where it generalises the ``electromagnetic duality'' to a more general class of quantum field theories. It is known that the Higgs branch $X$ and the Coulomb branch $X^{!}$ of such a theory are related in many non-trivial ways. For instance, as it was first observed in \cite{Oko18}, certain elliptic cohomology classes of $X$ and $X^{!}$, known as  the elliptic stable envelopes, are connected by a set of non-trivial relations.

In \cite{RSVZ21} we proposed to use these relations as a \textit{definition} of the $3D$-mirror symmetry. Speaking informally, our definition says that two varieties $X$ and $X^{!}$ are  $3D$-mirrors of each other if their elliptic stable envelope classes are related by a certain set of axioms.

The elliptic stable envelopes were defined in \cite{AO21}, see also \cite{DN23} for a  construction in the context of quantum field theory. The existence of these characteristic classes imposes certain conditions on varieties. Nevertheless, there are large classes of varieties for which the elliptic stable envelope exists, including the Nakajima varieties, the bow varieties, etc. Many examples of $3D$-mirror pairs $X,X^{!}$ satisfying the axioms of \cite{RSVZ21} has been recently constructed \cite{RSVZ21,RW,SZ22,BR23}. 

The set up of this paper is the following: let us assume that we are given a pair of varieties $X$ and $X^{!}$ which are related by the $3D$-mirror symmetry so that their elliptic stable envelope classes satisfy the axioms of \cite{RSVZ21}. We show that in this situation there exists a certain natural system of $q$-difference equations. In this paper we propose an explicit construction of these equations from the elliptic stable envelope classes.  It is expected that the difference equations we construct in this way are the quantum difference equations for $X$ and $X^{!}$ which govern the quantum K-theories of these varieties. 
The quantum difference equations (QDE) are the generalizations of the quantum differential equations in quantum cohomology to which they specialize in the limit $q\to 1$, see for instance \cite{TZ24}.  In this way, our construction allows one to extract the enumerative invariants of a variety, such as its quantum K-theory or cohomology, from the elliptic stable envelope classes.

The main idea of our construction is to use the elliptic stable envelope to describe the monodromy of QDE as it was suggested in \cite{AO21}. The monodromy of QDE is rigid in the sense that the QDE itself can be reconstructed from its monodromy.  

We start the exposition from a basic example of a scalar $q$-difference equation in the next section. In this example it is elementary to see that the monodromy of QDE is described by the elliptic stable envelope. It is also easy to see that the QDE can be reconstructed from the monodromy  via an elementary limiting procedure. This note can be considered as a generalization of this example to the higher rank situation. 

Finally, let us note that the QDE for Nakajima quiver varieties were constructed in \cite{OS22} using a different method - in \cite{OS22} for a Nakajima variety we fist construct a quantum group which acts on its equivariant K-theory. The building blocks of the QDE are then described in the representation theoretic terms, for instance, the wall-crossing operators for QDE are identified with the dynamical reflection operators in the corresponding quantum Weyl groups, see also \cite{TZ23}, for details of this construction in type $A$. In this note we skip this intermediate step and describe the QDE without any reference to the representation theory and quantum groups.

This is an overview paper based on the author's talk at the workshop {\it The 16th MSJ-SI: Elliptic Integrable Systems, Representation Theory 
and Hypergeometric Functions}, Tokyo 2023.  The narration is expository, we deal with all the relations informally and omit proofs. A more detailed discussion of ideas outlined here can be found in \cite{KS22,KS23,Ko,Sm21}.

\section*{Acknowledgements}
This work is supported by NSF grants DMS - 2054527, DMS-2401380 and by Simons Travel Support Grant for Mathematicians.  I would like to thank Hitoshi Konno for discussions and Tokyo University of Marine Science and Technology for hospitality.

\section{basic example \label{basexsec}} 
\subsection{} Let us consider the following $q$-difference equation (we assume $|q|<1$):
\bean \label{qdetp0}
\Psi(z q) = \Mop(z) \Psi(z), \ \ \ \textrm{where}  \ \ \ \Mop(z)=\dfrac{1-z }{1-z \hbar}.
\eean
This equation is the QDE for a ``point" Nakajima quiver variety $X=T^{*}\mathbb{P}^{0} \cong pt$.
The solution of this equation analytic near $z=0$ has the form:
$$
\Psi_{0}(z) = \sum\limits_{d=0}^{\infty}\, \dfrac{(\hbar)_{d}}{(q)_{d}} z^d = \dfrac{(z \hbar, q)_{\infty}}{(z,q)_{\infty}},
$$
where we denote
$$
(x)_d=(1-x)(1-x q)\dots (1-x q^{d-1}), \ \ \ (x,q)_{\infty}= \prod\limits_{i=0}^{\infty} (1-x q^{i}).
$$
The other solution is
$$
\Psi'_{\infty}(z) = e^{-\frac{\ln(z) \ln(h)}{\ln(q)}}\,  \Psi_{\infty}(z), \ \ \ \textrm{where}  \ \ \  \Psi_{\infty}(z) = \hbar^{1/2} \sum\limits_{d=0}^{\infty}\, \dfrac{(\hbar)_d}{(q)_{d}} \dfrac{q^d}{\hbar^d} z^{-d}= \hbar^{1/2} \dfrac{(q/z,q)_{\infty}}{(q/(z \hbar),q)_{\infty}}.
$$
Here $\Psi_{\infty}(z)$ is  part of the solution analytical near $z=\infty$. 
The transition ``matrix'' from the solution near $z=0$ to the solution near $z=\infty$ is called the monodromy of the $q$-difference equation:
\bean \label{monodtp0}
\Mon(z) = \Psi_{\infty}(z) \Psi_{0}(z)^{-1} =  \,\dfrac{\vartheta(z)}{ \vartheta(z \hbar)},
\eean
where 
\bean \label{thetdef}
\vartheta(z) = (z^{1/2}-z^{-1/2}) (q z)_{\infty} (q/z)_{\infty}
\eean
denotes the elliptic theta function. 
\subsection{}
The expression for monodromy (\ref{monodtp0}) has the following interpretation in terms of the $3D$-mirror variety. The $3D$-mirror of $X$ is the affine plane $X^{!} \cong \mathbb{C}^2$. 
It can be viewed as a symplectic variety equipped with an action of a torus $\torA^{!}=\mathbb{C}^{\times}_z\times \mathbb{C}^{\times}_\hbar$. The first factor acts on $X^{!}$  by $(x,y)\to (z x, z^{-1} y)$. Its action preserves the canonical symplectic form  $\omega=dx \wedge dy$ on $X^{!}$. The second acts by $(x,y)\to (x,\hbar^{-1} y)$, it scales the symplectic form with the weight $\hbar^{-1}$. 

The $\torA^{!}$-fixed set consists of a single point $(X^{!})^{\torA^{!}}=p$ corresponding to the origin of the affine plane.    Let us define the attracting sets of the fixed point by:
$$
\textsf{Attr}_{+}(p) = \{a \in X^{!}: \lim\limits_{z\to 0}\, z\cdot (a) =p\} = \{ y=0\},
$$
and 
$$
\textsf{Attr}_{-}(p) = \{a \in X^{!}: \lim\limits_{z\to \infty}\, z\cdot (a) =p\} = \{ x=0\}.
$$
In this case, the classes of the attracting sets $\Stab_{\pm}(p)=[\textsf{Attr}_{+}(p)]$ in the equivariant elliptic cohomology $\textrm{Ell}_{\torA^!}(X^{!})$ are known as the {\it elliptic stable envelopes} of $p$ corresponding to the chambers $z\to 0$ and $z\to \infty$. Restricting these classes to $p$ one finds:
\bean \label{stabstp0}
\left.\Stab_{+}(p)\right|_{p} = \vartheta(z) , \ \ \ \left.\Stab_{-}(p)\right|_{p}= \vartheta(z \hbar).
\eean 
The {\it elliptic $R$-matrix} of a variety $X^{!}$ is defined as the transition matrix between the stable envelopes of torus fixed points corresponding to opposite chambers \cite{AO21}. Therefore, from (\ref{stabstp0}) we find that in our example the $R$-matrix equals:
\bean \label{rmattp0}
\Rm_{X^{!}}(z) = \Stab_{-}^{-1} \circ \Stab_{+}  =  \dfrac{\vartheta(z)}{ \vartheta(z \hbar)}. 
\eean
This is our first important observation: {\bf the monodromy of the quantum difference equation of a variety $X$ coincides with the elliptic $R$-matrix of the $3D$-mirror variety $X^{!}$}.

\subsection{\label{monolimex}} 
One can ask is it possible to reconstruct the quantum difference equation (\ref{qdetp0}) from its monodromy (\ref{monodtp0})?  

For $s\in \mathbb{Q}$ (which we call ``slope'') we have the following limit for the monodromy (\ref{monodtp0}):
\bean\label{monolim1}
\lim\limits_{q \to 0}\, \Mon(z q^{s}) =\lim\limits_{q \to 0}\, \dfrac{\vartheta(z q^{s})}{ \vartheta(z \hbar q^{s})}  = \left\{\begin{array}{ll}
\hbar^{\lfloor s \rfloor +1/2}, \\
\dfrac{1-z }{ 1-z \hbar }\, \hbar^{s+1/2}.
\end{array}\right. 
\eean
where $\lfloor s \rfloor \in \mathbb{Z}$ denotes the floor function. 
This identity follows directly from our definition of the theta function  (\ref{thetdef}) as an infinite product - an elementary calculation shows that only  finitely many factors of this product contribute to the limit $q\to 0$ which gives (\ref{monolim1}).

Let $0<\epsilon\ll 1$ be an infinitely small positive real number, then from the previous limit we find:
\bean \label{correctedlim}
\Big( \lim\limits_{q \to 0}\, \Mon(z q^{s+\epsilon}) \Big)^{-1} \lim\limits_{q \to 0}\, \Mon(z q^{s}) = \left\{\begin{array}{cc}
1, & s \not \in \matZ,\\
\Mop(z), & s \in \matZ.
\end{array}\right.
\eean
A few things to note here: first, observe that (\ref{correctedlim}) is a piecewise constant function of $s\in \mathbb{Q}$ which ``jumps" at the ``hyperplane arrangement" in $\matQ$ corresponding to the integral points $\matZ\subset \mathbb{Q}$.    If $s$ is at a hyperplane, i.e. $z\in \matZ$ then  we obtain $\Mop(z)$ defining the QDE (\ref{qdetp0}). The last observation suggests that {\bf one can reconstruct the QDE by analysing the $q\to 0$ limit of the monodromy $\Mon(z q^{s})$ as a function of the slope $s$}.

\subsection{} 
Let us summarize the calculations we made for our basic example. First we showed that using elliptic cohomology of $3D$-mirror variety $X^{!}$ one can determine the monodromy of quantum difference equation for $X$.  Second, using certain limiting procedure, one can reconstruct the quantum difference equation from its monodromy. Speaking informally, this shows that the elliptic cohomology of the mirror  variety $X^{!}$ governs enumerative geometry of $X$. 

We will show that this example has a fairly straightforward generalization to the case of $X$ for which  the elliptic stable envelopes and $3D$-mirrors $X^{!}$ are defined. New features which appear for more general examples of $X$ are the following. First, in general, QDE (\ref{qdetp0}) is a first order $q$-difference equations of rank $r$, i.e., $\Mop(z)$ is a $r\times r$ matrix where $r=\textrm{rk}(K(X))$. The solution $\Psi(z)$, correspondingly, will be given by $r\times r$ matrix - the fundamental solution matrix of the QDE. 
Similarly, the elliptic R-matrices (\ref{rmattp0}) turn into rank $r$ matrices with elliptic matrix elements.

In general, there are several K\"ahler variables  $z=(z_1,\dots,z_m)$ for  some $m\in \matN$. Thus, the analog of the limit (\ref{monolim1}) is formulated for a slope parameter $s=(s_1,\dots,s_m) \in \matQ^{m}$. Similarly to (\ref{monolim1}) the $q\to0$ limit of the monodromy $\Mon(z q^s)$ for $z q^s=(z_1 q^{s_1},\dots,z_m  q^{s_m})$ turns out to be a piecewise constant function of $s$ which changes value only when $s$ crosses hyperplanes of a certain hyperplane arrangement $\frak{H}\subset \matQ^m$.

We shall identify the monodromy of QDE  for a variety $X$ with the elliptic $R$-matrix of $X^{!}$. 
We then reconstruct the QDE for $X$ by examining  the $q\to 0$ limit of the monodromy for non-generic values of the slope $s$, i.e., when $s\in \frak{H}$. In the last section, we sketch this construction for the case of the Hilbert scheme of points $X=\mathrm{Hilb}^{n}(\matC^2)$.

\section{QDE for Nakajima varieties} 
\subsection{\label{exsec}}
A  nice set of varieties for which the elliptic stable envelope exists is provided by the Nakajima varieties \cite{AO21}. These varieties are defined as GIT quotients
$$
X_{\theta} = T^{*} \textrm{Rep}(Q) /\!\!/_{\!\theta} \mathsf{G}  
$$
where $\textrm{Rep}(Q)$ is a complex vector space given by a representation of a quiver $Q$.  The parameter $\theta \in \mathbb{R}^{|Q|}$, where $|Q|$ is the number of vertices in the quiver, denotes a choice of the stability parameter for the GIT quotient. The space of stability parameters $\mathbb{R}^{|Q|}$ is equipped with a set of hyperplanes containing $0\in \mathbb{R}^{|Q|}$ which separate it into a set of chambers. The stability conditions $\theta$, $\theta'$ in the same chamber produce isomorphic GIT quotients $X_{\theta}\cong X_{\theta'}$.

Let $\kms$ be a complex toric variety, whose infinities correspond to chambers in $\mathbb{R}^{|Q|}$.  The non-equivalent Nakajima varieties, thus, correspond to the infinities of $\kms$, see Fig.\ref{stabchambs}.

\begin{figure}[H]
\begin{center}
\includegraphics[scale=0.3]{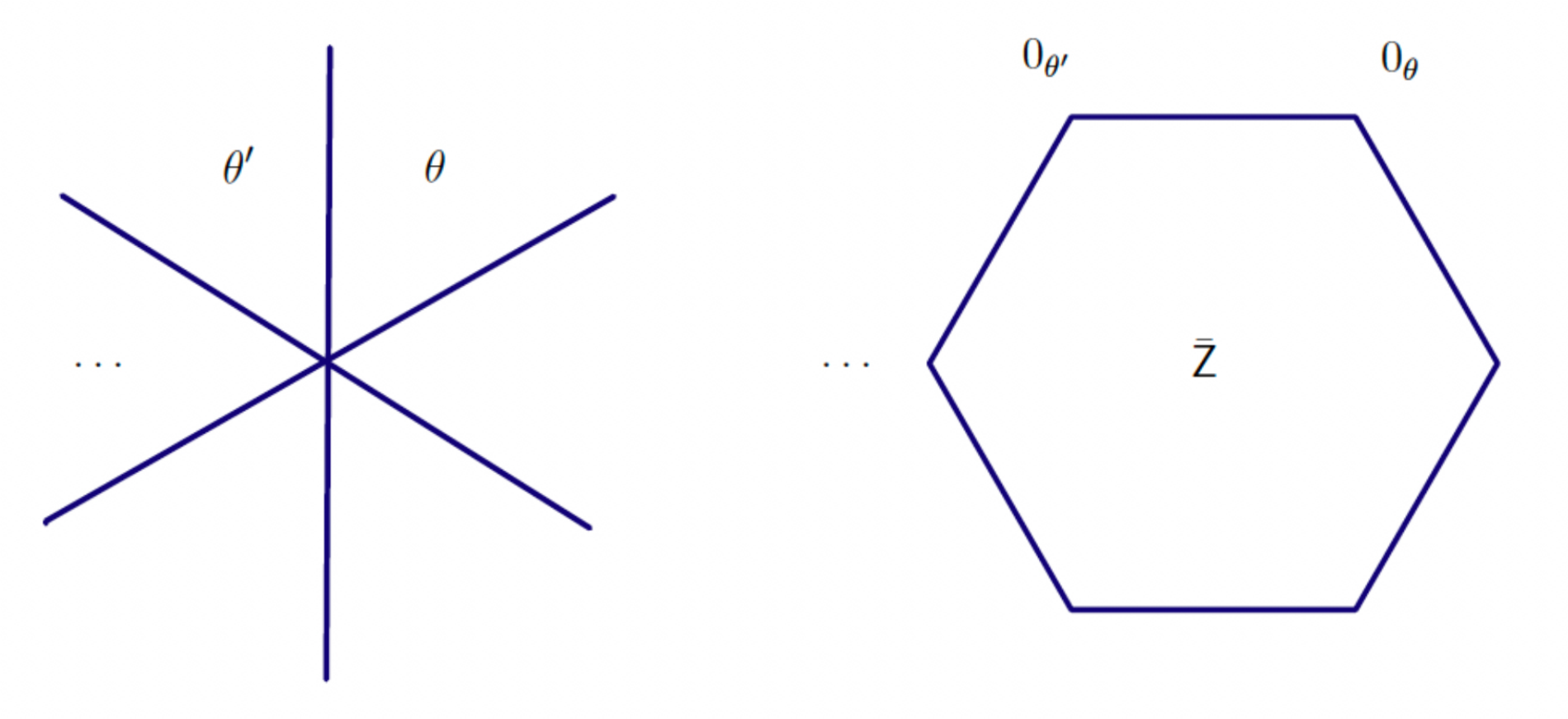}
\end{center}
\caption{Stability chambers and toric variety $
\bar{\mathsf{Z}}$ \label{stabchambs}} 
\end{figure}

\noindent
{\textbf{Example:}} Let us consider a quiver $Q$ consisting of one vertex and a framing Fig.\ref{grqv}. 
\begin{figure}[H]
\includegraphics[scale=0.3]{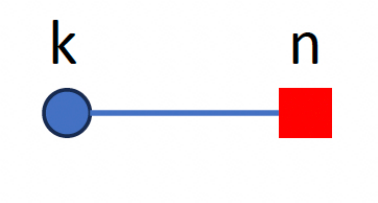}
\caption{The framed quiver representing cotangent bundles over Grassmannian \label{grqv}} 
\end{figure}
\noindent
Let us consider a representation  of $Q$ with dimension $k$ and framing dimension $n$. 
In this case, the space of the stability conditions is $\mathbb{R}$ and so there are only two chambers:
$$
\theta>0 , \ \ \ \theta<0.
$$
Computation shows that for a positive stability condition $X_{\theta}=T^{*}Gr(k,n)$ - the cotangent bundle over Grassmannian of $k$-subspaces in $\matC^n$. For a negative $\theta$ we obtain $X_{\theta}=T^{*} Gr(n-k,n)$, see Section 11 in \cite{MO19} for details. In this case the toric variety is $\bar{\mathsf{Z}} = \mathbb{P}^1$. The point $0\in \bar{\mathsf{Z}}$ corresponds to $X_{\theta}=T^{*}Gr(k,n)$ and $\infty\in \bar{\mathsf{Z}}$ corresponds to $X_{\theta}=T^{*}Gr(n-k,n)$.

\subsection{Counting curves in Nakajima varieties} 
The Nakajima  varieties have rich enumerative geometry, see \cite{Oko17} for an introduction. One of the most important elements of this theory is the ``{capping operator}'' which counts rational {\it quasimaps} to $X$ with non-singular and relative boundary conditions. We refer to Section 7.4 in \cite{Oko17} for the definition of the capping.  As a partition function of quasimaps, the capping is an element:
$$
\Psi_{X_\theta}(z,a)  \in K_{\torA}(X_{\theta})_{loc}^{\otimes 2} [[z]]
$$
If some basis of $K_{\torA}(X_{\theta})$ is fixed, then the capping is represented by a matrix:
\bean \label{cam}
\Psi_{X_\theta}(z,a)_{i,j} = \sum\limits_{H_2(X_{\theta},\matZ)_{\mathrm{eff}}} \, c_d(a)_{i,j}\, z^{d},\, \ \ \ i,j =1,\dots, \textrm{rk}(K_{\torA}(X_{\theta}))
\eean
where $z^{d}=z_1^{d_1}\cdots z_m^{d_m}$ denote ``K\"ahler parameters" counting the degrees $d= (d_1,\dots,d_m) \in H_2(X_{\theta},\matZ)$ of quasimaps.  
Speaking very informally, the coefficients of this generating function are computing degree $d$ curves in $X_{\theta}$: 
$$
c_{d}(a)_{i,j} =   \textrm{``number of degree $d$ rational curves in $X_{\theta}$ meeting classes $i$ and $j$'' }
$$
More precisely, the coefficients $c_{d}(a)_{i,j}$ represent certain $\torA$-equivariant integrals over the moduli space of quasimaps of degree $d$ with certain incidence conditions corresponding to $i,j \in K_{\torA}(X_{\theta})$. The equivariant integrals are computed by localization and thus are rational functions of the equivariant parameters $a$, i.e.,  $c_{d}(a)_{i,j} \in K_{\torA}(pt)_{loc} \cong \matQ(a)$.

The sum in (\ref{cam}) is over the cone of  effective curve classes $H_{2}(X,\matZ)_{\textrm{eff}}\subset H_{2}(X,\matZ) \cong \matZ^{|Q|}$ which is determined by the stability $\theta$.  It is natural to think of the K\"ahler parameters $z=(z_1,\dots, z_m)$ as local coordinates on $\kms$ near the infinity corresponding to $X_{\theta}$. Capping (\ref{cam})  is thus a certain {\bf analytic function in a neighborhood of $0_{\theta} \in \kms$}.

\subsection{}
It is known that $\Psi_{\theta}(z,a)$ is a fundamental solution matrix of a certain natural $q$-difference equation \cite{OS22}. This QDE has the following form:
\bean \label{qdegener}
\Psi_{X_\theta}(z q^{\lb},a)\lb = \Mop_{\lb}(z) \Psi_{X_\theta}(z,a), \ \ \lb \in \mathrm{Pic}(X_{\theta}) \cong \matZ^{|Q|}
\eean
An important conjecture states that the QDE {\it is independent of the stability condition $\theta$}.   This means that the capping operators $\Psi_{\theta}(z,a)$ for Nakajima varieties $X_{\theta}$ with various stability conditions $\theta$ provide a set of the fundamental solution matrices of {\it the same QDE.} The fundamental solution  $\Psi_{\theta}(z,a)$ corresponding a particular $\theta$ is distinguished by its analiticity near  $0_{\theta}\in \kms$. 

The natural question is to describe the transition matrix between two such fundamental solutions: 
$$
\Psi_{X_\theta}(z,a) = \Mon_{\theta \leftarrow \theta'}(z)\, \Psi_{X_{\theta'}}(z,a)
$$
The transition matrix $\Mon_{\theta \leftarrow \theta'}(z)$ is called {\it monodromy} of QDE. In practice, it is a certain rank $\textrm{rk}(K_{\torA}(X_{\theta}))$ matrix with elements given by elliptic functions of all parameters involved. 

\subsection{} 
The partition function satisfies also a certain dual system of $q$-difference equations in the equivariant parameters $a$:
\bean \label{shifteq}
\Psi_{X_\theta}(z,a q^{\sigma}) = \mathsf{S}_{\sigma}(z,a) \Psi_{X_\theta}(z,a), \ \ \sigma \in \textrm{cochar}(\torA) \cong \matZ^{\dim \torA},
\eean
which are called shift equations, see Section 8.2 in \cite{Oko17}. The dependence of $\Psi_{X_\theta}(z,a)$ on the K\"ahler parameters $z$ and the equivariant parameters $a$ is different, however.   $\Psi_{X_\theta}(z,a)$ is analytic in $z$ (i.e., is a power series in $z$ converging in some neighborhood of $0_{\theta}$). The coefficients $c_{d}(a)$ are rational functions of $a$ and thus $\Psi_{X_\theta}(z,a)$  is not an analytic in ~$a$. Nevertheless, with the notion of $3D$-mirror symmetry the K\"ahler parameters and the equivariant parameters enter the story on equal terms as we will see in the next section. 

\section{Monodromy of QDE from $3D$-mirror variety}
\subsection{}  The symmetry $a\longleftrightarrow z$ is explained best in the language of the so called {\it $3D$-mirror symmetry}. The $3D$-mirror symmetry suggests the following picture:  for  varieties $X_{\theta}$ there exists a set of companion varieties $X^{!}_{\sigma}$ labeled by $\sigma \in \mathsf{cochar}(\torA)$.  Each such variety is equipped with a natural equivariant curve counting which provides the capping operators:
$$
\Psi_{X^{!}_{\sigma}}(a,z)= \sum\limits_{d \in H_{2}(X^{!}_{\sigma},\matZ)_{\textrm{eff}}}\, c_{d}(a^{!}) \, (z^!)^{d} 
$$
where we denote by $a^!$  the equivariant parameters  and by  $z^!$  the K\"ahler parameters of $X^!_{\sigma}$.
One axiom of the  $3D$-mirror symmetry is the identification of variables
\bean 
\label{keequiv}
z = a^{!}, \ \ \ a = z^{!}
\eean
i.e., the equivariant parameters of $X_{\theta}$ are identified with the K\"ahler parameters on the mirror side $X^{!}_{\sigma}$ and vice versa. The second axiom is the existence of an isomorphism of the  equivariant K-theories of $X_\theta$ and $X^{!}_{\sigma}$. The main point is: under all these identifications the capping operators $\Psi_{X^{!}_{\sigma}}(a^!,z^!)$ provide fundamental solution matrices for {\bf the same system of $q$-difference equations} (\ref{qdegener}) and (\ref{shifteq}). The fundamental solutions $\Psi_{X^{!}_{\sigma}}(a^!,z^!)$ are distinguished by their analyticity in $z^{!}$, i.e., via identification (\ref{keequiv}), they are analytic solutions in the equivariant parameters $a$ (in a certain region). 

In summary, the capping operators of $X_{\theta}$ provide fundamental solution matrices analytic in $z$  and the capping operators of the mirror varieties $X^{!}_{\sigma}$ provide solutions analytic in $a$. This cures the asymmetry between the K\"ahler and the equivariant parameters discussed at the end of the previous section.

\subsection{} 
Similarly to the case of $X_{\theta}$, the non-equivalent choices of mirror varieties $X^{!}_{\sigma}$ are labeled by infinities of a certain toric variety $\ems$ obtained by a compactification of the equivariant torus $\torA$.  The structure of $\ems$ can be seen from the equivariant geometry of $X_{\theta}$ as follows. For a choice of $\sigma \in \mathsf{cochar}(\torA)$ and a torus fixed point component $i \in X^{\torA}_{\sigma}$ we define the attracting set:
\bean \label{attrset}
\mathsf{Attr}_{\sigma}(i) = \{x \in X_{\sigma}: \lim\limits_{a\to 0}\, \sigma(a) \cdot x \in i \}.
\eean
The space $\mathsf{cochar}(\torA) \otimes \mathbb{R} \cong \mathbb{R}^{\dim A}$ splits into a union of cones: two $\sigma$ and $\sigma'$ are in the same cone if they produce the same attracting sets: $\mathsf{Attr}_{\sigma}(i)=\mathsf{Attr}_{\sigma'}(i)$ for all $i\in X^{\torA}_{\theta}$.

This structure can also be seen as follows. Let us consider the union of hyperplanes:
\bean 
\coprod\limits_{\alpha}\, \{ \langle \alpha, \sigma\rangle =0 \}\, \subset  \mathsf{cochar}(\torA) \otimes \mathbb{R} \cong \mathbb{R}^{\dim \torA},
\eean
where $\alpha$ runs over the set of $\torA$-characters appearing in the normal bundles $N_{i}(X_{\theta})$ to all torus fixed components $i \in X_{\theta}^{\torA}$. These hyperplanes partition $\mathbb{R}^{\dim \torA}$ to the union of cones we need.  The toric variety $\ems$ is a compactification of the equivariant torus $\torA$ obtained by adding infinity points $0_{\sigma}$ corresponding to these cones.

The whole picture is now symmetric: there are two toric varieties $\kms$ and $\ems$, with local coordinates given by the K\"ahler and the equivariant parameters respectively. The QDE (\ref{qdegener})-(\ref{shifteq}) is invariant under the toric coordinate changes on $\kms$ and on $\ems$. In other words the QDE for all $X_{\theta}$ and all $X^{!}_{\sigma}$ are equivalent. The corresponding capping operators are distinguished by the following properties: $\Psi_{X_{\theta}}(z,a)$ is the unique fundamental solution matrix which is analytic in the neighborhood of $0_{\theta}\in  \kms$, similarly $\Psi_{X^{!}_{\sigma}}(z,a)$ is the unique fundamental solution regular near $0_{\sigma} \in \ems$, see Fig.\ref{monodpics}. 
\subsection{} 
Let us fix two infinity points $0_{\theta}\in \kms$ and $0_{\sigma} \in  \ems$. These points define two fundamental solution matrices  $\Psi_{X_{\theta}}(z,a)$ and $\Psi_{X^{!}_{\sigma}}(z,a)$. Let $ \zamon_{\theta \leftarrow \sigma}(z,a)$ denote the  transition matrix between these two bases of solutions
\bean \label{ellmonod}
\Psi_{X^{!}_{\sigma}}(z,a) = \zamon_{\sigma \leftarrow \theta}(z,a)  \Psi_{X_{\theta}}(z,a).
\eean
Note that $\Psi_{X^{!}_{\sigma}}(z,a)$ is analytic  in the neighborhood of $0_{\sigma} \in  \ems$. It is non-analytic in $z$ and typically has infinitely many poles  $z$ in any neighborhood of $0_{\theta}\in \kms$. For $\Psi_{X_{\theta}}(z,a)$ the situation is opposite - it is analytic near $0_{\theta}\in \kms$ and has  poles accumulating in any neighborhood of $0_{\sigma} \in  \ems$. The matrix $\Sigma_{\sigma \leftarrow \theta}(a,z)$, thus, has an interesting property: it cancels the poles in $a$ and adds new poles in $z$. For this reason the matrix $\Sigma_{\sigma \leftarrow \theta}(a,z)$ is sometimes refereed to as the ``pole subtraction matrix''. 

Such pole subtraction matrices can be constructed geometrically \cite{AO21} using the stable envelope classes in the equivariant elliptic cohomology. Speaking very informally, these are the elliptic cohomology classes of the attracting sets: 
\bean \label{ellstabs}
\Stab^{X_{\theta}}_{\sigma} (i) = [\textsf{Attr}_{\sigma}(i)] \in \textrm{Ell}_{\torA}(X_{\theta}).
\eean
Precise definition of these classes is more sophisticated and we refer to \cite{AO21} for details.
The corresponding transition matrices have the form:
\bean \label{ellmattrans}
\zamon_{\sigma \leftarrow \theta}(a,z)_{i,j} = \left.\Stab^{X_{\theta}}_{\sigma} ({i})\right|_{j},  \ \ \ i,j \in (X_{\theta})^{\torA}.
\eean
The class (\ref{ellstabs}) is called {\it the elliptic stable envelope} of the fixed point component $i\in X^{\torA}_{\theta}$. In terms of the elliptic stable envelope maps we can write:
$$
\zamon_{\sigma \leftarrow \theta}(a,z) = \Big(\Stab^{X_{\theta}}_{\sigma}\Big)^{t}
$$
where $t$ denotes the transposition. 

\subsection{} 
Since the $3D$-mirror varieties come to our picture on equivalent grounds, the relation (\ref{ellmonod}) remains true if we exchange $X_{\theta}$ and $X^{!}_{\sigma}$, i.e., we have a transition matrix:
\bean \label{invtrans}
\Psi_{X_{\theta}}(z,a)= \zamon_{\theta \leftarrow \sigma}(a,z) \Psi_{X_{\sigma}^{!}}(z,a) 
\eean
which is given by the elliptic stable envelope classes of the $3D$-mirror variety $X^{!}_{\sigma}$
$$
\zamon_{\theta \leftarrow \sigma}(a,z) = \Big(\Stab^{X^{!}_{\sigma}}_{\theta}\Big)^{t}.
$$
Note that by (\ref{keequiv}) we may identify the stability condition $\theta$ of $X_{\theta}$ with a cocharacter of a torus $\torZ$ acting on $X^{!}_{\sigma}$. Using this cocharacter we can define the attracting sets and the elliptic envelope classes for the mirror $X^{!}_{\sigma}$ in the same way as in (\ref{attrset})  and (\ref{ellstabs}).

We note that by construction:
\bean \label{invells}
\zamon_{\sigma \leftarrow \theta}(a,z)= \zamon_{\theta \leftarrow \sigma}(a,z)^{-1}
\eean
It is known that the inverse of the stable envelope matrix is obtained by taking the opposite cocharacter $\sigma\to -\sigma$ and  a transposition,  Proposition 3.4 in \cite{AO21}. Thus, in components the equality (\ref{invells}) takes the form of a matrix identity:
\bean \label{mirsymells}
\left.\Stab^{X_{\theta}}_{\sigma} (i)\right|_{j} =  \left.\Stab^{X^{!}_{\sigma}}_{-\theta} (j)\right|_{i}.
\eean
In (\ref{mirsymells}) we assume that the parameters on both sides are identified via (\ref{keequiv}), and $X^{\torA}_{\theta}=(X_{\sigma}^{!})^{\torZ}$ via 3D-mirror symmetry. 

Equation (\ref{mirsymells}) is remarkable: it relates the elliptic tautological classes of different varieties.  One existing approach postulates (\ref{mirsymells}) as the {\bf definition of $3D$-mirror symmetry.} We may define that $X_{\theta}$ and $X^{!}_{\sigma}$ form  a $3D$-mirror pair, if there exists an identification of parameters such that (\ref{mirsymells}) holds. This approach was first suggested in \cite{RSVZ21} and tested in the example of $X=T^{*} Gr(k,n)$ - cotangent bundles to Grammarians (see example of Section \ref{exsec} above). Since then, more examples of $3D$-mirror pairs satisfying this condition were constructed, see \cite{RSVZ21,SZ22,RW}. A very rich set  of $3D$-mirror pairs has been recently constructed using type $A$ - bow varieties \cite{BR23}. Type $A$ bow varieties are naturally closed with respect to $3D$-mirror symmetry, i.e., $3D$-mirror of a bow variety is also a bow variety. For this reason, describing the QDE for the bow varieties using the methods we outline in the present note might  be an interesting  problem.


\subsection{} 
Let $0_{\theta}, 0_{-\theta} \in \kms$ be two opposite infinities. Let $\Psi_{X_{\theta}}(z,a)$ and $\Psi_{X_{-\theta}}(z,a)$ be the analytic solutions of QDE near these points. We denote by $\Mon_{\theta}(z,a)$ the transition matrix between these bases of solutions
$$
\Psi_{X_{\theta}}(z,a) = \Mon_{\theta}(z,a) \Psi_{X_{-\theta}}(z,a).
$$
The matrix $\Mon_{\theta}(z,a)$ is called the monodromy of QDE associated with $X_{\theta}$. From the discussion of the previous sections we have:
\bean \label{monodrfactor}
\Mon_{\theta}(z,a) =\zamon_{\sigma \leftarrow -\theta}(z,a)^{-1} 
 \circ \zamon_{\sigma \leftarrow  \theta}(z,a) 
\eean
for any choice of $\sigma$ see Fig.\ref{monodpics}. By (\ref{ellmattrans}) we have
\bean \label{moviastab}
\Mon_{\theta}(z,a) = \Big((\Stab^{X_{-\theta}}_{\sigma})^{t}\Big)^{-1} \circ  \Big(\Stab^{X_{\theta}}_{\sigma} \Big)^{t},
\eean
and by (\ref{mirsymells}) we obtain
\bean \label{monodviastab}
\Mon_{\theta}(z,a) = \Big(\Stab^{X^{!}_{\sigma}}_{\theta}\Big)^{-1} \circ  \Stab^{X^{!}_{\sigma}}_{-\theta}.
\eean
The last formula expresses the monodromy matrix as the transition matrix from the basis of the elliptic cohomology of $X^{!}_{\sigma}$ given by the elliptic stable envelope classes with cocharacter $\theta$ to the basis of the elliptic stable envelope classes with the cocharacter $-\theta$. In  \cite{AO21} this transition matrix was called {\it the elliptic $R$-matrix} of $X^{!}_{\sigma}$.  

\begin{figure}[H]
\includegraphics[scale=0.23]{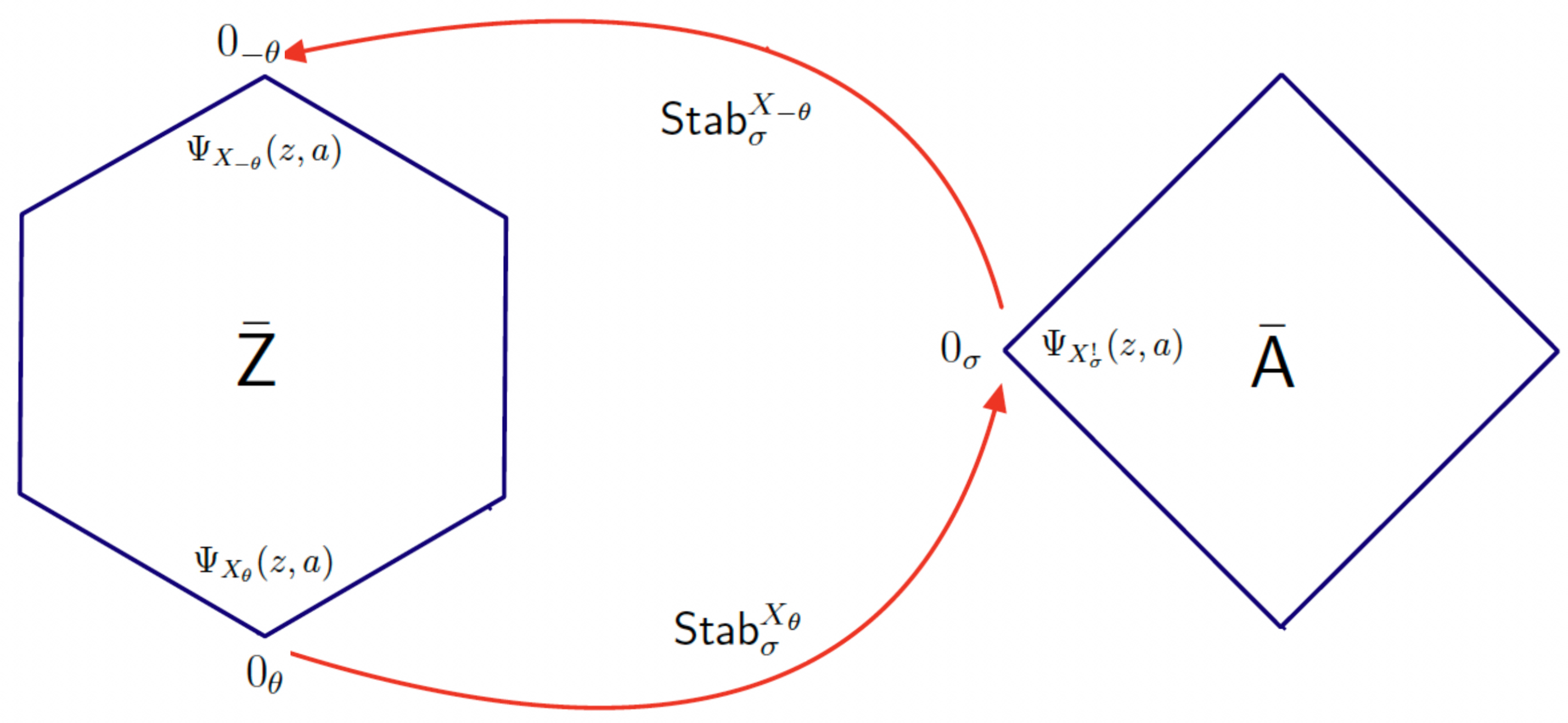}
\caption{Transition matrices between various fundamental solutions \label{monodpics}} 
\end{figure}
\noindent

In summary,  the monodromy of QDE can be determined via computation in the equivariant elliptic cohomology. In the next section we discuss a limiting procedure which allows one to reconstruct the QDE from its monodromy.

\section{K-theoretic limit of the monodromy matrix}
\subsection{Factorization of K-theoretic limits}  As in Section \ref{monolimex} we would like to investigate the limit of the monodromy matrix
\bean 
\label{monolim}
\lim\limits_{q\to 0}\, \Mon(z q^{s},a), \ \ \ \textrm{where} \ \ \  z q^{s} = (z_1 q^{s_1},\dots, z_m q^{s_m}),
\eean
as a function of  the slope $s=(s_1,\dots, s_m) \in H^{2}(X_{\theta},\mathbb{Q}) \cong \mathbb{Q}^m$. By (\ref{moviastab})   it is sufficient to understand this limit for the elliptic stable envelope. We now review the results of \cite{KS22,KS23,Ko} where this analysis was performed under assumption that $X^{\torA}_{\theta}$ is finite, see also \cite{Oko21} for a less restrictive approach.  
Let us consider the stable envelope matrix $\zamon_{\sigma \leftarrow \theta}(a,z)$
 defied by (\ref{ellmattrans}). Recall that the space of the slopes $\mathbb{Q}^m$ is equipped with a $\textrm{Pic}(X_{\theta}) \cong \matZ^{m}$ - periodic hyperplane arrangement $\frak{H}$ \cite{OS22}. If $s$ is generic, i.e., $s\not \in \frak{H}$, then  Proposition 4.3 in \cite{AO21} gives:
\bean \label{genlim}
\lim\limits_{q\to 0}\, \zamon_{\sigma \leftarrow \theta}(a,z q^s) = \mathcal{A}^{X_{\theta},s}_{\sigma}(a)
\eean
where $\mathcal{A}^{X_{\theta},s}_{\sigma}(a)$ is the matrix of the fixed point components of the K-theoretic stable envelope classes (see Section 2 of \cite{OS22} for definition of K-theoretic stable envelope) with slope~$s$:
\bean \label{kthstmat}
\mathcal{A}^{X_{\theta}, s}_{\sigma}(a)_{i,j} = \left.\Stab^{Kth,\,X_{\theta},\,  s}_{\sigma}(i)\right|_{j} \in K_{\torA}(pt), \ \ \ i,j \in X_{\theta}^{\torA}.
\eean
Components of this matrix take values in the equivariant K-theory of a point, i.e., they are Laurent polynomials in the equivariant parameters $a$. Note that the matrix  $\mathcal{A}^{X_{\theta}, s}_{\sigma}(a)$ does not depend on the K\"ahler variables $z$. It is known that $\mathcal{A}^{X_{\theta}, s}_{\sigma}(a)$ is a piecewise constant function of the slope $s$ which changes value only when $s$ crosses a hyperplane of $\frak{H}$ (compare this behavior with (\ref{monolim1}) in our basic example).

The most interesting to us is the case of non-generic slope $s \in \frak{H}$.  For such slopes the limit will depend on both the equivariant $a$ and the K\"ahler parameters $z$ (again, compare this behavior with (\ref{monolim1})). The most interesting feature of this limit, is that the dependence on $a$ and $z$ can be separated: the limit factors to a product of two matrices one of which depends only on $a$ and the second, essentially only on $z$ \cite{KS23}.

Let $s$ be an arbitrary slope, possibly at a wall or at an intersection of several hyperplanes of $\frak{H}$. Let $s\pm \epsilon $ be generic slopes, obtained from $s$ by a small shift in the direction $\pm \theta$. Then, the limit with slope $s$ can be factored (Theorem 3 in \cite{KS23}):
\bean \label{factorizat}
\lim\limits_{q\to 0}\, \zamon_{\sigma\leftarrow\, \theta}(a,z q^s) =  \widetilde{\mathcal{Z}}^{s,\pm }_{\theta}(z) \mathcal{A}_{\sigma}^{X_{\theta},s\pm \epsilon}(a).
\eean
The second factor in the right side is the matrix of the K-theoretic stable envelope (\ref{kthstmat}) corresponding to the generic slopes parameters $s+ \epsilon$ or $s-\epsilon$.

The matrix $\widetilde{\mathcal{Z}}^{s,\pm }_{\theta}(z)$ has the following form
\bean \label{Zvonj}
\widetilde{\mathcal{Z}}^{s,\pm }_{\theta}(z) = \cL^{s} \, \mathcal{Z}^{s,\pm}_{\theta}(z) \, (\cL^{s})^{-1}
\eean
where in the right side we denote by  $\cL^s \in \Pic(X_{\theta}) \otimes \matQ$  the fractional line bundle corresponding to $s$, namely $c_1(\cL^{s})=s \in H^{2}(X_{\theta},\matQ)$. In the basis of torus fixed points $\cL^s$ is represented by a diagonal matrix with eigenvalues given by certain monomials in equivariant parameters $a$ with fractional powers $s$, see (\ref{frakO}) for an example.

Finally, the matrix $\mathcal{Z}^{s,\pm}_{\theta}(z)$ {\it depends only on the K\"ahler variables $z$} and has the following geometric meaning. For a slope $s$ we may associate certain finite group $G_s \subset \mathsf{Z}$ acting on the mirror variety $X^{!}_{\sigma}$ \cite{KS23}.  Let $X^{!}_s = (X^{!}_{\sigma})^{G_s} \subset X^{!}_{\sigma}$ be the subvariety in the $3D$-mirror variety fixed by $G_s$.
The matrix $\mathcal{Z}^{s,\pm }_{\theta}(z)$ then has the following description:
\bean  \label{Zmats}
\mathcal{Z}^{s,\pm }_{\theta}(z)_{i,j} = \left.\Stab^{X^{!}_{s},Kth,\, \pm \epsilon}_{\theta}(i)\right|_{j} , \ \ \ i,j \in (X^{!}_s)^{\torZ},
\eean
i.e., this is the matrix of the K-theoretic stable envelope for $X^{!}_s$ with small ample or anti-ample slopes $\pm \epsilon$. Note that since $\torZ$ is commutative, these varieties have the same torus fixed points: $(X^{!}_s)^{\torZ}= (X^{!}_{\sigma})^{\torZ}$. 

We note that if $s$ is generic, then $X^{!}_s = (X^{!}_{\sigma})^{\torZ}$ is a finite set of points. In this case $\mathcal{Z}^{s,\pm}_{\theta}(z)=1$ and (\ref{factorizat}) is same as  (\ref{genlim}).

Note that (\ref{factorizat}) produces two different factorization corresponding two choices of the slopes $s+\epsilon$ and $s-\epsilon$. 
Once again, in above formulas $\epsilon$ denotes a small ample slope for $X_{\theta}$. One has to be careful when using this: for the variety with the opposite stability condition $X_{-\theta}$ the small ample slope is $-\epsilon$ and  the analog of the formula (\ref{factorizat}) in this case is:
\bean \label{factorizat2}
\lim\limits_{q\to 0}\, \zamon_{\sigma\leftarrow\, -\theta}(a,z q^s) =  \widetilde{\mathcal{Z}}^{s,\mp }_{-\theta}(z)\, \mathcal{A}_{\sigma}^{X_{-\theta},s\pm \epsilon}(a).
\eean
\subsection{Limiting factorizations} 
Note that if before taking the limit in (\ref{factorizat}) we in addition deform the argument $z\to z q^{\epsilon}$ by a small ample $\epsilon$ then the right side turns to $\mathcal{A}_{\sigma}^{X_{\theta},s\pm \epsilon}(a)$ by (\ref{genlim}). We note also that 
 $z q^{\epsilon} \to 0_{\theta} \in \bar{\torZ}$ as $q\to 0$. Thus from (\ref{factorizat}) we obtain:
 \bean \label{Zlimit1}
 \widetilde{\mathcal{Z}}^{s,+}_{\theta}(0_{\theta}) \mathcal{A}_{\sigma}^{X_{\theta},s+ \epsilon}(a) =\widetilde{\mathcal{Z}}^{s,-}_{\theta}(0_{\theta}) \mathcal{A}_{\sigma}^{X_{\theta},s- \epsilon}(a) =\mathcal{A}_{\sigma}^{X_{\theta},s+ \epsilon}(a), 
 \eean 
in particular, 
\bean
\widetilde{\mathcal{Z}}^{s,+}_{\theta}(0_{\theta}) =1.
\eean
Same logic for shifts $z \to q^{-\epsilon}$ gives:
 \bean 
 \widetilde{\mathcal{Z}}^{s,+}_{\theta}(0_{-\theta}) \mathcal{A}_{\sigma}^{X_{\theta},s+ \epsilon}(a) =\widetilde{\mathcal{Z}}^{s,-}_{\theta}(0_{-\theta}) \mathcal{A}_{\sigma}^{X_{\theta},s- \epsilon}(a) =\mathcal{A}_{\sigma}^{X_{\theta},s- \epsilon}(a), 
 \eean
 in particular, 
\bean
\widetilde{\mathcal{Z}}^{s,-}_{\theta}(0_{-\theta}) =1.
\eean
Applying same argument to (\ref{factorizat2}) we also obtain:
\bean
\widetilde{\mathcal{Z}}^{s,\mp }_{-\theta}(0_{-\theta})\, \mathcal{A}_{\sigma}^{X_{-\theta},s\pm \epsilon}(a)=\mathcal{A}_{\sigma}^{X_{-\theta},s- \epsilon}(a), \\
\widetilde{\mathcal{Z}}^{s,\mp }_{-\theta}(0_{\theta})\, \mathcal{A}_{\sigma}^{X_{-\theta},s\pm \epsilon}(a)=\mathcal{A}_{\sigma}^{X_{-\theta},s+ \epsilon}(a).
\eean 
In particular
\bean \label{Zlimit2}
\widetilde{\mathcal{Z}}^{s,+ }_{-\theta}(0_{-\theta})= \widetilde{\mathcal{Z}}^{s,- }_{-\theta}(0_{\theta})=1.
\eean

\subsection{Limits of the monodromy}
From  (\ref{monodrfactor})  and using (\ref{factorizat}) with slope $-$ and (\ref{factorizat2}) with slope $+$ we find:
\bean \label{limnongen}
\lim\limits_{q \to 0}\, \Mon(z q^{s})=\mathcal{A}_{\sigma}^{X_{-\theta},s- \epsilon}(a)^{-1}  \cL^{s} \,  \mathcal{Z}^{s,+}_{-\theta}(z)^{-1}  \mathcal{Z}^{s,+}_{\theta}(z) \, (\cL^{s})^{-1}  \mathcal{A}_{\sigma}^{X_{\theta},s+ \epsilon}(a)
\eean
The middle factor here $\mathcal{R}^{X^!_s}(z)=\mathcal{Z}^{s,+}_{-\theta}(z)^{-1}  \mathcal{Z}^{s,+}_{\theta}(z)$ is the transition matrix between two K-theoretic stable bases in K-theory of $X^{!}_{s}$, corresponding to cocharacters $\theta$ and $-\theta$. In terminology of \cite{OS22} the matrix  $\mathcal{R}^{X^!_s}(z)$ is the  K-theoretic R-matrix of $X^!_{s}$. 
For generic slopes $s\not\in\frak{H}$, $X^!_{s}$ is a finite set, $\mathcal{Z}^{s,\pm}_{\theta}(z)=\mathcal{Z}^{s,+}_{-\theta}(z)=1$
and therefore:
\bean \label{limgen}
\lim\limits_{q \to 0}\, \Mon(z q^{s})=\mathcal{A}_{\sigma}^{X_{-\theta},s}(a)^{-1}  \mathcal{A}_{\sigma}^{X_{\theta},s}(a),  \ \ \ \textrm{if}  \ \ \  s\not\in\frak{H}.
\eean
\subsection{The wall crossing operators}
As in Section \ref{basexsec}, formula (\ref{correctedlim}) we define  two types of {\bf wall-crossing} operators:
\bean \label{wcdefs}
 \ \ \ \ \wc_{s}(z)=\Big(\lim\limits_{q \to 0}\, \Mon(z q^{s-\epsilon})\Big)^{-1} \, \lim\limits_{q \to 0}\, \Mon(z q^{s}), \ \ \wc^{*}_{s}(z)= \lim\limits_{q \to 0}\, \Mon(z q^{s})\, \Big(\lim\limits_{q \to 0}\, \Mon(z q^{s+\epsilon})\Big)^{-1}
\eean
We call $\wc_{s}(z)$ the wall-crossing operator for $X_{\theta}$ and $\wc^{*}_{s}(z)$
the wall-crossing operator for  $X_{-\theta}$ corresponding to $s$. 
We note that (\ref{wcdefs}) defines the wall crossing operators for an arbitrary slope $s\in H^{2}(X_{\theta},\matQ)$, including the case when $s$ belongs to an intersection of several hyperplanes of $s\in \frak{H}$. We note also that for generic slopes
\bean \label{regularwc}
\wc_{s}(z) =\wc^{*}_{s}(z) =1 , \ \ \ \textrm{if} \ \ \ s\not \in \frak{H}.
\eean
Form (\ref{limnongen}) and (\ref{limgen}) we obtain:
$$
\wc_{s}(z)= \left\{\begin{array}{ll}
1, & s\not \in \frak{H},\\
\mathcal{A}_{\sigma}^{X_{\theta},s- \epsilon}(a)^{-1}  \widetilde{\mathcal{R}}^{X^!_s}(z)  \mathcal{A}_{\sigma}^{X_{\theta},s+ \epsilon}(a), & s\in \frak{H}.
\end{array}\right.
$$
and 
$$
\wc^{*}_{s}(z)= \left\{\begin{array}{ll}
1, & s\not \in \frak{H},\\
\mathcal{A}_{\sigma}^{X_{-\theta},s- \epsilon}(a)^{-1}  \widetilde{\mathcal{R}}^{X^!_s}(z)  \mathcal{A}_{\sigma}^{X_{-\theta},s+ \epsilon}(a), & s\in \frak{H}.
\end{array}\right.
$$
where we denote by $\widetilde{\mathcal{R}}^{X^!_s}(z)=\cL^{s} \,  \mathcal{R}^{X^!_s}(z) \, (\cL^{s})^{-1}$ the K-theoretic $R$-matrix of $X^!_s$ twisted by the fractional line bundle $\cL^{s}$.

The last formulas say that  the wall-crossing operator $\wc_{s}(z)$ acts in the stable basis of $K_{\torA}(X_{\theta})$ as the K-theoretic $R$-matrix of $X^{!}_s$. To be more precise, it says that applying the operator $\wc_{s}(z)$ to the stable basis $\Stab^{s-\epsilon}_{\sigma}(i)$ and expanding the results in the stable basis $\Stab^{s+\epsilon}_{\sigma}(j)$, $i,j\in (X_{\theta})^{\torA}$ we obtain the matrix which coincides with the twisted R-matrix of $X^{!}_s$:
\bean \label{actioninstab}
\langle \Stab^{X_{\theta},s+\epsilon}_{\sigma}(i) |  \wc_{s}(z) | \Stab^{X_{\theta},s-\epsilon}_{\sigma}(j) \rangle = \widetilde{\mathcal{R}}^{X^!_s}(z)_{i,j}.
\eean
Similarly, $\wc^{*}_{s}(z)$ acts naturally in $K_{\torA}(X_{-\theta})$ by:
\bean 
\langle \Stab^{X_{-\theta},s+\epsilon}_{\sigma}(i) |  \wc^{*}_{s}(z) | \Stab^{X_{-\theta},s-\epsilon}_{\sigma}(j) \rangle = \widetilde{\mathcal{R}}^{X^!_s}(z)_{i,j}.
\eean
This formula gives a tool for computing the action of the wall-crossing operators on $K_{\torA}(X_{\theta})$ if the $3D$-mirror variety is known.

\subsection{The wall-crossing operators at infinities of $\bar{\torZ}$}
Combining (\ref{limnongen})-(\ref{wcdefs}), from formulas (\ref{Zlimit1})-(\ref{Zlimit2}) we get: 
\bean \label{bthatzero}
\wc_{s}(0_{\theta}) = \wc^{*}_{s}(0_{-\theta}) =1
\eean 
i.e., the wall crossing operators for $X_{\theta}$ and $X_{-\theta}$ trivialize in at the infinities of $\bar{\torZ}$ corresponding to these varieties. 
\subsection{Transport operators}
We denote limit (\ref{limgen})  by: 
$$
{\bf T}^{s}: =\mathcal{A}_{\sigma}^{X_{-\theta},s} (a)^{-1} \mathcal{A}_{\sigma}^{X_{\theta},s} (a), \ \ \ s\not\in\frak{H}
$$
then from (\ref{wcdefs}) we find
\bean \label{BtoBbar}
\wc^{*}_{s}(z)\, {\bf T}^{s+\epsilon}={\bf T}^{s-\epsilon}\, \wc_{s}(z).
\eean
Thus from (\ref{BtoBbar}) and (\ref{bthatzero}) we also obtain:
\bean \label{wcrelat}
\wc^{*}_{s}(0_{\theta})\, {\bf T}^{s+\epsilon}={\bf T}^{s-\epsilon}, \ \ \ {\bf T}^{s+\epsilon}={\bf T}^{s-\epsilon}\, \wc_{s}(0_{-\theta}).
\eean
Let $s$ and $s'$ be two generic slopes. Let us fix a path from $s$ to $s'$ which intersects $\frak{H}$ at the points $w_1,w_2,\dots w_m$ in the positive direction. Iterating (\ref{wcrelat}) several times we obtain:
\bean \label{recurwc}
\wc^{*}_{w_m}(0_{\theta})\dots \wc^{*}_{w_1}(0_{\theta})\, {\bf T}^{s} = {\bf T}^{s'}, \ \ \  \wc_{w_m}(0_{-\theta})\dots \wc_{w_1}(0_{-\theta})\, {\bf T}^{s'} = {\bf T}^{s}.
\eean
We call the  ${\bf T}^{s}$ {\it the transport operator} with slope $s$. It can be shown that this operator computes the transport (or monodromy) of the Dubrovin connection for $X_{\theta}$ from $0_{\theta}$ to $0_{-\theta}$ over the line with slope $s$, see Section 8 in \cite{Sm21}. 

From the ``window'' condition for the $K$-theoretic stable envelope \cite{OS22} it follows that the wall-crossing operators  satisfy:
 \bean \label{conjwall}
 \cL^{-1} \wc_{s}(z) \cL= \wc_{s-\cL}(z),  \ \   \cL^{-1} \wc^{*}_{s}(z) \cL= \wc^{*}_{s-\cL}(z), \ \ \  \mathcal{L} \in \Pic(X_{\theta}).
 \eean

\section{Reconstructing QDE from monodromy matrix \label{reconsec}}
\subsection{$q$-difference equations} Let $\mathcal{L} \in \Pic(X_{\theta})$ be a line bundle. Let $-\epsilon \in H^{2}(X_{\theta}, \matQ)$ be a small generic anti-ample slope and $-\epsilon - \cL$ be the slope shifted by $\cL$. 
 Choose a path from $-\epsilon$ to  $-\epsilon - \cL$ which intersects the hyperplane arrangement $\frak{H}$ at $w_1,\dots, w_m$. We define operators
 \bean \label{moper}
 \Mop_{\cL}(z)=\cL \, \wc_{w_m}(z q^{-w_m}) \dots \wc_{w_1}(z q^{-w_1}). 
 \eean
and 
\bean 
\Mop^{*}_{\cL}(z)=\cL \, \wc^{*}_{w_m}(z q^{-w_m}) \dots \wc^{*}_{w_1}(z q^{-w_1}). 
\eean
where $q\in \matC^{\times}$ is a complex parameter. We will assume that $|q|<1$.

We call $\Mop_{\cL}(z)$ and $\Mop^{*}_{\cL}(z)$ the \textrm{$q$-difference operators} for $X_{\theta}$ and $X_{-\theta}$ respectively. It follows from our construction of the wall-crossing operators that $\Mop_{\cL}(z)$ and $\Mop^{*}_{\cL}(z)$ are well-defined, i.e., the are independent of the choice of a path from $-\epsilon$ to  $-\epsilon - \cL$.

In particular, we have
 $$
  \Mop_{\cL_1 \cL_2}(z) = \Mop_{\cL_1}(z q^{\cL_2})   \Mop_{\cL_2}(z) =   \Mop_{\cL_2}(z q^{\cL_1})  \Mop_{\cL_1}(z).
 $$
 and the same for $\Mop^{*}_{\cL_1}(z)$.
The last equation means that the following systems of $q$-difference equations 
 \bean \label{qdeofx}
 \Psi(z q^{\cL}) \cL = \Mop_{\cL}(z) \Psi(z), \ \ \ \cL \in \Pic(X_{\theta}),
 \eean
 and 
 \bean  \label{diff2}
 \Psi(z q^{\cL}) \cL = \Mop^{*}_{\cL}(z) \Psi(z), 
 \eean
 are well defined.  From (\ref{BtoBbar}) we have
\bean \label{transpM}
({\bf T}^{-\epsilon})^{-1}\, \Mop^{*}_{\cL}(z)\, {\bf T}^{-\epsilon} =  \Mop_{\cL}(z).
\eean
and thus (\ref{qdeofx}) and (\ref{diff2}) are equivalent systems of $q$-difference equations. In the next section we show that (\ref{qdeofx}) and (\ref{diff2}) are exactly the QDEs for $X_{\theta}$ and $X_{-\theta}$. 

Since by (\ref{bthatzero}) $\Mop_{\cL}(0_{\theta})$ is diagonal in the basis of torus fixed points, the columns of the matrix $\mathbf{T}^{-\epsilon}$ are the eigenvectors of $ \Mop^{*}_{\cL}(0_{\theta})$. Similarly,  $(\mathbf{T}^{-\epsilon})^{-1}$ gives the eigenvectors of $\Mop_{\cL}(0_{-\theta})$. 
 \subsection{Monodromy of $q$-difference equation}
The $q$-difference equation (\ref{qdeofx}) is constructed so that its monodromy coincides with monodromy of QDE (\ref{moviastab}). Since $q$-difference equations with the same monodromy are equivalent, it follows that (\ref{qdeofx}) and (\ref{diff2}) are the QDEs for $X_{\theta}$ and $X_{-\theta}$ respectively.

To see it,  we note that by the properly (\ref{conjwall}) and by $\Pic(X_{\theta})$-invariance of the hyperplane arrangement $\frak{H}\subset\mathbb{Q}^m$ the infinite product
\bean \label{infprof0}
 \Psi_{\theta}(z) = \wc_{w_{-1}}(z q^{-w_{-1}})^{-1} \wc_{w_{-2}}(z q^{-w_{-2}})^{-1} \cdots=\prod\limits_{i< 0}^{\leftarrow}\, \wc_{w_{i}}(z q^{{-w_{i}}})^{-1} 
\eean
is a formal solution (\ref{qdeofx}). The product in (\ref{infprof0}) runs over a set of intersections with hyperplanes $w_i$ of a path from a small anti-ample slope $-\epsilon$ to infinity in the direction of $-\theta$, see Fig.\ref{pathandmon}b). By the arrow in the product we indicate the the operators are multiplied in the order  in which $\langle s,w_i\rangle$ increases: $\langle s,w_{-1}\rangle\geq \langle s,w_{-2}\rangle \geq \dots$.   By (\ref{bthatzero})  this infinite product converges for  $z$ in a  small analytic neighborhood of~$0_{\theta} \in \kms$. Thus (\ref{bthatzero}) is a fundamental solution matrix of (\ref{qdeofx}) analytic near~$0_{\theta}$.

Similarly, by (\ref{conjwall}) the infinite product 
\bean \label{secondsolutinf}
 \Psi_{-\theta}(z) = \wc^{*}_{w_{0}}(z q^{-w_{0}}) \wc^{*}_{w_{1}}(z q^{-w_{1}}) \cdots=\prod\limits_{i\geq 0}^{\rightarrow}\, \wc^{*}_{w_{i}}(z q^{{-w_{i}}}) \,
\eean 
is a formal solution of (\ref{diff2}). 
The product in (\ref{secondsolutinf}) runs over a set of hyperplanes intersected at $w_i$ when one follows over a path from a small anti-ample slope $-\epsilon$ to infinity in the direction of $\theta$, see Fig.\ref{pathandmon}a). By (\ref{bthatzero}) this infinite product converges in the neighborhood of $0_{-\theta}$. Thus (\ref{secondsolutinf}) is the fundamental solution matrix of (\ref{diff2}) analytic near~$0_{-\theta}$. 

By (\ref{transpM}) $({\bf T}^{-\epsilon})^{-1}\,  \Psi_{-\theta}(z)$ is also a solution of (\ref{qdeofx}) near $z=0_{-\theta}$. Thus, the monodromy of (\ref{qdeofx}) is the product of wall crossing operators over ``all walls'' in the hyperplane arrangement Fig.\ref{pathandmon}c).
\bean \label{infmonodt}
\Mon(z,a)=\Psi_{\theta}(z)^{-1} \Psi_{-\theta}(z)=\Big(\prod\limits_{i<0}^{\rightarrow} \, \wc_{w_i}(z q^{-w_i})\Big) \, ({\bf T}^{-\epsilon})^{-1} \, \Big(\prod\limits_{i\geq 0}^{\rightarrow}\,  \wc^{*}_{w_{i}}(z q^{{-w_{i}}})\Big) . 
\eean
We constructed the solutions as infinite products in full analogy with our basic example considered in Section~\ref{basexsec}. In particular, the infinite product (\ref{infmonodt}) is a higher rank analog of (\ref{monodtp0}) which is also an infinite product by (\ref{thetdef}). In this way, the monodromy (\ref{infmonodt}) may be thought of as a ``matrix theta function''. 

\subsection{Limits of fundamental solutions}
Assume that $s$ is on the line passing through $0\in H^{2}(X_{\theta},\matQ)$ in the direction of $\theta$. Then 
$$
\lim\limits_{q\to 0} q^{s}\to \left\{\begin{array}{ll} 0_{\theta} , & \langle s,\theta \rangle >0 \\
0_{-\theta}, & \langle s,\theta \rangle <0 \\
1, & s=0
\end{array}\right. 
$$
Thus, from (\ref{infprof0}) and (\ref{bthatzero}) we find:
\bean \label{limitpsi1}
\lim\limits_{q\to 0} \Psi_{\theta}(z q^{s})=\left\{\begin{array}{ll}
1, &  \langle s,\theta \rangle \geq 0,\\
\\ & \\
\Big(\wc_{w_{-1}}(0_{-\theta})^{-1}\wc_{w_{-2}}(0_{-\theta})^{-1}\dots \Big) \wc_{s}(z)^{-1} & \langle s,\theta \rangle < 0.
\end{array}\right. 
\eean
Similarly from (\ref{secondsolutinf}) we obtain:
\bean \label{limitpsi2}
\lim\limits_{q\to 0} \Psi_{-\theta}(z q^s) = \left\{\begin{array}{ll}
\Big(\wc^{*}_{w_0}(0_{\theta})\wc^{*}_{w_1}(0_{\theta}) \dots\Big) \wc^{*}_{s}(z) ,& \langle s, \theta \rangle \geq 0 ,\\
1,& \langle s, \theta \rangle <  0
\end{array}\right.
\eean
In (\ref{limitpsi1}) and (\ref{limitpsi2})  $\wc_{s}(z)=\wc^{*}_{s}(z)=1$ when $s\not \in \frak{H}$ by (\ref{regularwc}).

\begin{figure}[H]
\begin{center}
\includegraphics[scale=0.55]{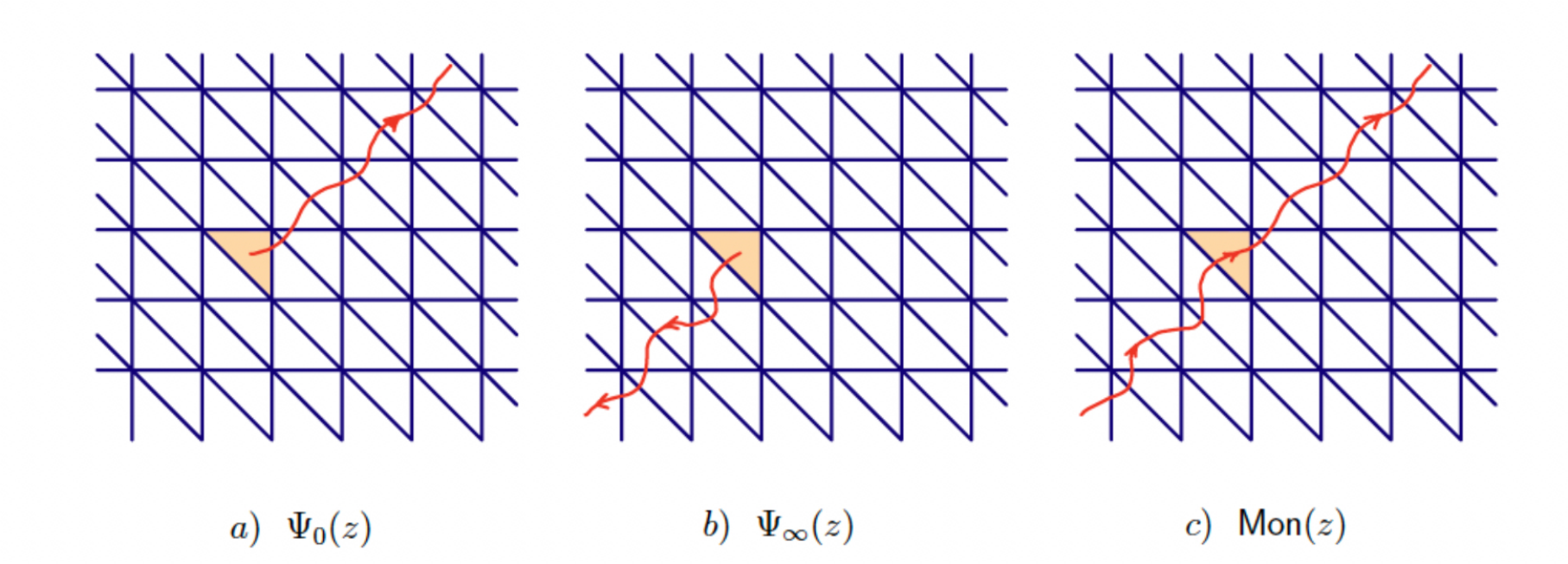}
\end{center}
\caption{Fundamental solutions and the monodromy matrices as ordered infinite products  of the wall-crossing operators \label{pathandmon}} 
\end{figure}
Combining all this together, from (\ref{infmonodt}) we obtain:
$$
\lim\limits_{q\to 0}\, \Mon(z q^{s},a) = 
\left\{\begin{array}{ll}
({\bf T}^{-\epsilon})^{-1} \cdot \Big(\wc^{*}_{w_0}(0_{\theta})\wc^{*}_{w_1}(0_{\theta}) \dots\Big) \wc^{*}_{s}(z) , & \langle s,\theta \rangle \geq 0, \\
\\
\wc_{s}(z) \, \Big(\dots  \wc_{w_{-2}}(0_{-\theta}) \wc_{w_{-1}}(0_{-\theta}) \Big)   \cdot ({\bf T}^{-\epsilon})^{-1}, & \langle s,\theta \rangle <0.
\end{array}\right. 
$$
Using (\ref{recurwc}) we can also write it as:
$$
\lim\limits_{q\to 0}\, \Mon(z q^{s},a) = 
\left\{\begin{array}{ll}
({\bf T}^{s-\epsilon})^{-1} \wc^{*}_{s}(z)  , & s\geq 0, \\
\\
  \wc_{s}(z) ({\bf T}^{s+\epsilon})^{-1}  , & s<0.
\end{array}\right. 
$$
and thus by (\ref{BtoBbar}) we see that for an arbitrary $s$ we may write:
$$
\lim\limits_{q\to 0}\, \Mon(z q^{s},a) ={\bf T}^{s-\epsilon} \, \wc_{s}(z)  = \wc^{*}_{s}(z)\, {\bf T}^{s+\epsilon}.
$$
For a generic slope $s\not \in \frak{H}$:
$$
\lim\limits_{q\to 0}\, \Mon(z q^{s\pm \epsilon},a) = {\bf T}^{s\pm\epsilon} , 
$$
and therefore
$$
\Big(\lim\limits_{q\to 0}\, \Mon(z q^{s-\epsilon},a) \Big)^{-1} \lim\limits_{q\to 0}\, \Mon(z q^{s},a) = \wc_{s}(z),
$$

$$
 \lim\limits_{q\to 0}\, \Mon(z q^{s},a)\, \Big(\lim\limits_{q\to 0}\, \Mon(z q^{s+\epsilon},a) \Big)^{-1} = \wc^{*}_{s}(z).
$$
Thus, we checked that the limit $q\to0$ of the monodromy of (\ref{qdeofx}), (\ref{diff2}) coincides with the same limit for the monodromy of QDE for all  $s$. It follows that the monodromies coincide.

\section{An example: the QDE for the Hilbert scheme $X_{\theta}=\mathrm{Hilb}^{n}(\matC^2)$. \label{sechilbex}}

\subsection{The hyperplane arrangement}
In this section we outline the construction of the QDE in an example of the quiver variety given by the Hilbert scheme $X_{\theta}=\mathrm{Hilb}^{n}(\matC^2)$. A detailed exposition of this example may be found in \cite{Sm21}. 

We identify $H^{2}(X_{\theta}, \matQ) \cong \matQ$. The first step is to determine the hyperplane arrangement in this space using limit (\ref{factorizat}). This limit is a piecewise constant function of $s\in \matQ$ and the hyperplanes corresponds to the values of $s$ where (\ref{factorizat}) is discontinuous.

A combinatorial formula for the elliptic stable envelope of the Hilbert scheme $X_{\theta}$ was obtained in \cite{Sm20}. Using this formula one can analyze (\ref{factorizat}) and determine the hyperplane arrangement $\frak{H}$, see Theorem 9 in \cite{Sm20}:
\bean \label{wallset}
\frak{H}= \{s=\dfrac{a}{b} \in \matQ: 0<|b|\leq n \} \subset \matQ. 
\eean

\subsection{The wall crossing operators}
The Hilbert scheme $X_{\theta}$ is self-dual with respect to the $3D$-mirror symmetry $X^{!}_{\sigma} \cong \mathrm{Hilb}^{n}(\matC^2)$ \cite{KZ23}. Given a wall $s=\frac{a}{b}$ from the set (\ref{wallset}), we consider the group of $b$-th roots of unity
$$
G_s = \langle e^{2 \pi  i s} \rangle  \subset \torZ \cong \matC^{\times}.
$$
Where $\torZ$ is the torus acting on the mirror Hilbert scheme $X^{!}_{\sigma}$. The fixed points $(X^{!}_{\sigma})^{G_s}$ are determined from the quiver description of $X^{!}_{\sigma}$. As a quiver variety, $X^{!}_{\sigma}$ is described by the Jordan quiver Fig.\ref{jordpic}a), with $\matC^n$ assigned to the vertex and $\matC$ assigned to the framing vertex.

The group $G_s$ acts on $X^{!}_{\sigma}$  via loop rotation, i.e., by $(I,J,A,B)\to (I,J,e^{2 \pi  i s} A,e^{-2 \pi  i s} B)$. By the standard argument,  the vector space $\matC^n$ as a representation of $G_s$ splits to the weight subspaces
$
\matC^n = \matC^n_{0}\oplus \dots \oplus \matC^n_{b-1}
$
so that an element $e^{-2 \pi  i s} \in G_s$ acts on $\matC^n_{m}$ via multiplication by  $e^{2 \pi m i s}$. Thus, the $G_s$-invariant matrix elements of the operators $A=\oplus_i A_i,B= \oplus_i B_i$ are:
$$
A_i : \matC^n_{i} \to  \matC^n_{i+1},  \ \ \ B_i : \matC^n_{i+1} \to  \matC^n_{i}.
$$
and the components of the $G_s$-fixed points in $X^{!}_{\sigma}$ are described by the cyclic quiver varieties with $b$ -vertices Fig.\ref{jordpic}b):
\bean \label{cyclcomponent}
X^{!}_{s}:=(X^{!}_{\sigma})^{G_s} = \coprod\limits_{n_0+\dots+n_{b-1}=n} X(n_0,\dots, n_{b-1}).
\eean
\begin{figure}[H]
\begin{center}
\includegraphics[scale=0.25]{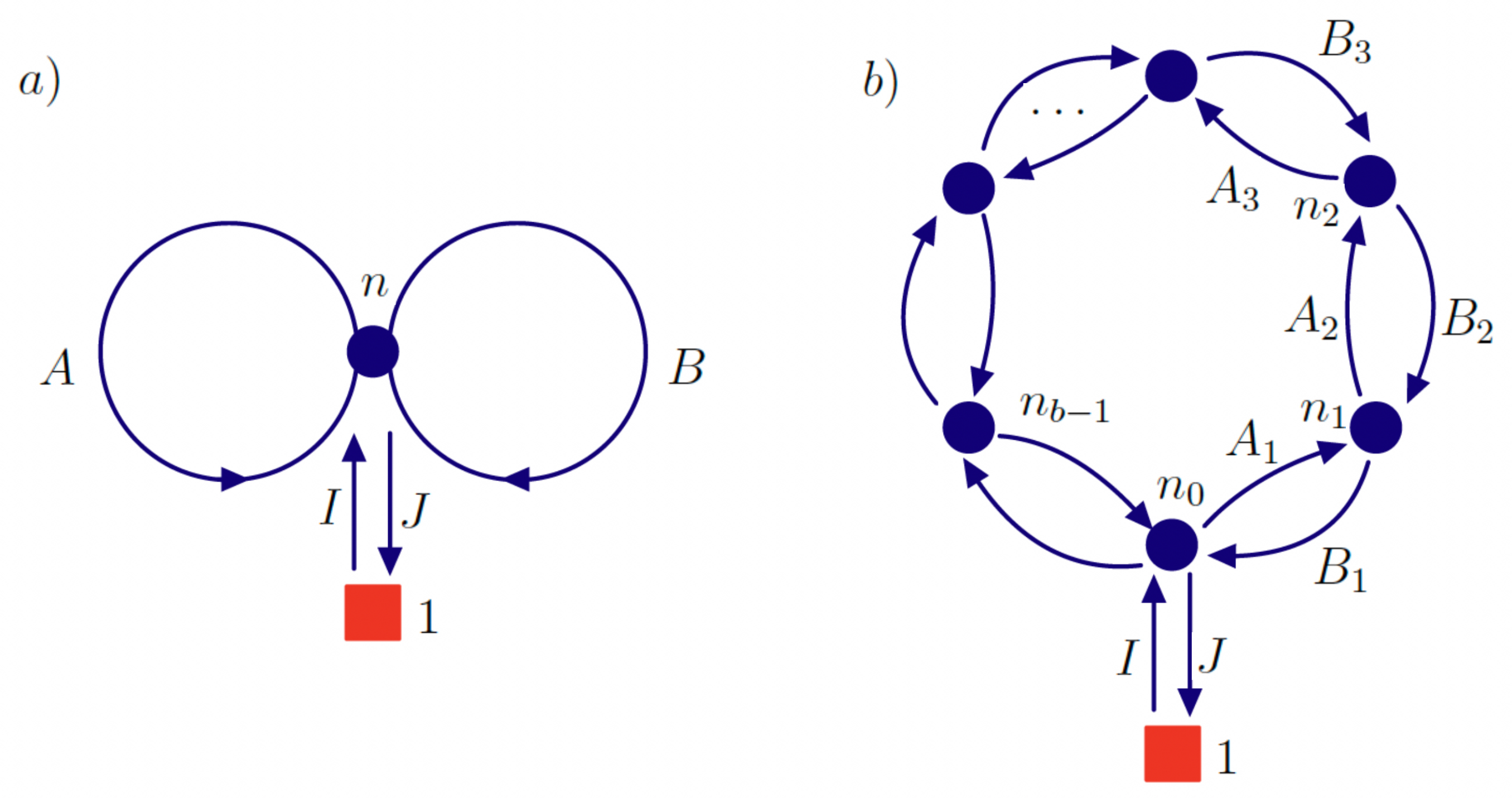}
\end{center}
\caption{a) Framed Jordan quiver b) Framed cyclic quiver with $b$ vertices \label{jordpic}} 
\end{figure}
Let $\mathcal{Z}^{s,- }_{\theta}(z)$ be the matrices of K-theoretic stable envelopes (\ref{Zmats}) for the cyclic varieties (\ref{cyclcomponent}) for small anti-ample slope. The matrices $\mathcal{Z}^{s,- }_{\theta}(z)$ can be computed using  combinatorial formulas of \cite{Din23}.  
Let $\mathcal{R}^{X^!_s}(z)$ be the R-matrix of the cyclic quiver variety:
$$
\mathcal{R}^{X^!_s}(z)=\mathcal{Z}^{s,-}_{-\theta}(z)  \mathcal{Z}^{s,-}_{\theta}(z)^{-1}
$$
and let $\widetilde{\mathcal{R}}^{X^!_s}(z) = \mathcal{O}(1)^{s}  \mathcal{R}^{X^!_s}(z) \mathcal{O}(1)^{-s}$ be its twist by the fractional line bundle $\mathcal{O}(1)^{s}$ corresponding to $s$.  Explicitly,  in the basis of torus fixed points, which for the Hilbert schemes $\textrm{Hilb}^{n}(\matC^2)$ are labeled by the partitions of $n$, it is given by a diagonal matrix with the following eigenvalues:
\bean \label{frakO}
\mathcal{O}(1)^{s} = \textrm{diag}\Big( \left.\mathcal{O}(1)^{s}\right|_{\lambda}: \lambda \in \{\textrm{partitions of} \  n \} \Big), \ \ \ \left.\mathcal{O}(1)^{s}\right|_{\lambda} = \prod\limits_{(i,j)\in \lambda}\, t_1^{s (i-1)} t_2^{s (j-1)}.
\eean 
We note that despite the weights  $\left.\mathcal{O}(1)^{s}\right|_{\lambda}$ for $s\in \matQ$ are fractional, the matrix elements of the twisted $R$-matrix $\widetilde{\mathcal{R}}^{X^!_s}(z)$ are integral, see Theorem 3 in \cite{KS23}. Finally, equation (\ref{actioninstab}) defines the action of the wall-crossing operator $\wc_{w}(z)$ in the stable basis of K-theory $K_{\torA}(X_{\theta})$.
\subsection{}
The ``hyperplane arrangement'' (\ref{wallset}) is clearly $\Pic(X_{\theta}) \cong \matZ$ - invariant. The generator $\mathcal{O}(1) \in \Pic(X_{\theta})$ acts on   $H^{2}(X_{\theta},\matQ)$ by 
$w\to w+1$. The anti-canonical slopes correspond the alcove $(-1/n,0)\subset \matQ$. Thus, the procedure of Section  \ref{reconsec}  gives the following formula for the operator (\ref{moper}):
$$
 \Mop_{\mathcal{O}(1)}(z) = \mathcal{O}(1) \prod\limits_{{w \in \frak{H}}:\atop {-1 \leq w<0}}\, \wc_{w}(z),
$$
where in the right side we denote by the same symbol $\mathcal{O}(1)$ the operator acting on $K_{\torA}(X)$ via multiplication by the line bundle $\mathcal{O}(1)$. For example, for the Hilbert scheme of  $n=4$ points   this operator is given by the following product of the wall-crossing operators:
$$
 \Mop_{\mathcal{O}(1)}(z) = \mathcal{O}(1) \wc_{-1}(z) \wc_{-3/4}(z) \wc_{-2/3}(z) \wc_{-1/2}(z) \wc_{-1/3}(z) \wc_{-1/4}(z) .
$$
The quantum difference equation for the Hilbert scheme of points has the form:
\bean \label{qdehilb}
\Psi(z q) \mathcal{O}(1) =  \Mop_{\mathcal{O}(1)}(z) \Psi(z).
\eean
The fundamental solution matrix of this equation normalized  by 
$\Psi(0) = \mathcal{O}_{X_{\theta}}$ 
provides the  capping operator for the Hilbert scheme, see Section 7.4 \cite{Oko17}. The capping operator plays fundamental role in the  DT theory of threefolds, see for instance  \cite{KOO}. We note also that in the cohomological limit, the $q$-difference equation turns into differential equation   discovered and studied in \cite{OP1,OP2}. We refer  to \cite{Sm21} where this story is discussed in greater details. 

Finally, an interested reader may compare the results of this section with Section 7 of \cite{OS22}, where the quantum difference equation for the Hilbert scheme $\mathrm{Hilb}^{n}(\matC^2)$ was obtain using representation theoretic approach. In particular, in \cite{OS22} we expressed the wall-crossing operators $\wc_w(z)$ in terms of the toroidal quantum group $\frak{gl}_1$ acting on $K_{\torA}(X_{\theta})$. This description is quite different from  the geometric constructions of this note, see also \cite{TZ23} for representation theoretic approach for the cyclic quivers.

\end{document}